# Overestimated isomer depletion due to contamination

Song Guo[1,2], Yongde Fang[1,2], Xiaohong Zhou[1,2] & C. M. Petrache[3]

The recent paper by Chiara et al.[1] provided the first experimental evidence of nuclear excitation by electron capture (NEEC), responding a long-standing theoretical prediction. NEEC was inferred to be the main channel to excite an isomer in Molybdenum-93 to a higher state, leading to a rapid release of full isomer energy (isomer depletion). The deduced large excitation probability $P_{exc}$=0.010(3) for this mechanism implied strong influence on the survival of nuclei in stellar environments. However, the excitation probability is much higher than the estimated NEEC probability $P_{NEEC}$ according to a following theoretical work[2] by approximately 9 orders of magnitude. Nevertheless, the reported $P_{exc}$ is predicted to be due to other unknown mechanism causing isomer depletion, which is expected to open up a new era of the storage and release of nuclear energy. Here we report an analysis of the reported experimental results, showing that the observed isomer depletion is significantly overestimated due to the contamination.

A 268-keV γ-ray emission is used to mark the NEEC process. Since it is mainly populated directly from the fusion evaporation reaction, the real NEEC events were identified by the coincidence with the transitions above the 2.4 MeV isomer. In a gamma rich environment, considerable false events due to the contamination cannot be excluded event by event according to the time coincidence and the gamma-ray energy. Here the contamination means events not induced by isomer depletion, which mainly consists of the chance coincidences, coincidences with Compton background, coincidences with other transitions with similar energies, and so on. It is treated based on two prerequisites which are too idealistic in the Methods section of Ref. 1. First, the chance coincidences are considered negligible. Second, only the smooth Compton background is taken into account.

According to the spectra in Extended Data Fig. 3 reported in the supplementary information of Ref. 1, the existence of the 263-keV peaks indicates that the chance coincidences are not negligible. Emitted from the $T_{1/2}$ = 6.85 h isomer, this transition is in coincidence only with the 685- and 1,478-keV transitions while no γ-ray with higher energy is emitted. According to the first prerequisite, no 263-keV peak should be found in coincidence with the 2,475-keV transition, which is in contradiction to the observation. From a rough estimation based on the reported experimental setup (see the supplementary information), it also shows that the chance coincidences are not negligible.

About the second prerequisite, the Compton background is approximately smooth only in a certain range. For example, there is usually a bump near 200 keV in the Compton background which cannot be regarded as smooth, while a Doppler-shifted 241-keV peak was used to deduce the excitation probability in Ref. 1. Besides the Compton background, there can be some other possible mechanisms such as an unknown weak transition with a Doppler-shifted energy range overlapped with that of the gating transition. Considering these effects, it cannot be preconceived that the ratios between backgrounds in coincidence with the peak energy range and that of the nearby channels of the gating transition ($k$ in the Method section of Ref. 1), are same among different transitions.

[1]*Key Laboratory of High Precision Nuclear Spectroscopy and Center for Nuclear Matter Science, Institute of Modern Physics, Chinese Academy of Sciences, Lanzhou 730000, People's Republic of China*
[2]*School of Nuclear Science and Technology, University of Chinese Academy of Science, Beijing 100049, People's Republic of China*
[3]*Centre de Sciences Nucléaires et Sciences de la Matière, CNRS/IN2P3, Université Paris-Saclay, Bât. 104-108, 91405 Orsay, France*
✉*e-mail: gs@impcas.ac.cn*

Due to the idealistic prerequisites, it is possible that the contamination was not properly excluded. In Fig. 3b of Ref. 1, the 263-keV transition was clearly seen as a peak, and marked as a known transition in $^{93}$Mo. There is a known 262-keV transition above the gating transitions[3], but it should be excluded since a Doppler-shifted transition could not appear as a peak in a spectrum without Doppler correction. Therefore, the only known transition in $^{93}$Mo at this energy is the one below the isomer which is expected to be absent in this spectrum if the contamination is properly excluded. Furthermore, the intensity of this peak is about twice that at 268 keV, which is similar with the ratio in the spectrum gated by 1,478-keV transition (see Fig. 2b in Ref. 1). Considering that only the 268-keV transition contributes to the isomer depletion, the similar ratios indicate that this peak in the spectrum is mainly from the contamination. Therefore, the deduced excitation probability $P$ = 0.010(3) should be regarded as an upper limit of the isomer depletion.

In contrast, the peak at 263 keV did vanish in the spectrum doubly gated by 1,478- and 2,475-keV transitions (see Fig. 2a in Ref. 1), which was regarded as the proof of the existence of isomer depletion. However, it can be explained without invoking the isomer depletion mechanism. As mentioned before, the 263-keV peak cannot be found in a spectrum gated by the 2,475-keV transition due to true events. However, the situation for the 268-keV transition is quite different since it was mainly populated via a series of transitions involving high-energy statistical γ-rays. The background induced by the high-energy γ-rays would cover the gating range of the Doppler-shifted 2,475-keV transition, allowing the existence of 268-keV peak in this spectrum due to true events. Therefore, comparing to the 268-keV transition, the 263-keV one is significantly shrunk in Fig. 2a in Ref. 1. Partly populated from the isomer, the 685-keV transition is less affected. The intensities of the 123-, 203-, 770- and 963-keV transitions are much lower, and it is not surprising that they are unidentifiable in the spectrum in Fig. 2a in Ref. 1 due to the poor statistics. Furthermore, the Fig. 2a was probably obtained using improper background parameters (see the supplementary information).

Since the reported excitation probability is only an upper limit, it should be measured again, possibly with a setup that can assure to fully exclude the contamination. A possible refined method is delivering the residues to another location by adding an extra beam line. With a beam line with a length of several meters, the prompt γ-rays are left behind and there will be only the isomer depletion which can give rise to the observation of some special γ-ray as the 268-keV transition in $^{93}$Mo. The time coincidence between the γ-rays and the stopped residues can further help to distinguish it from the spontaneous decay. Comparing the yields of the isomer depletion and the spontaneous decay from the isomer, it would be easy to deduce the excitation probability. By this method, it is possible to measure it with a precision of $10^{-4}$ - $10^{-6}$. Proper stopping material is expected to maximize the probability of NEEC, which is possible with a realistic theoretical estimation. Such a measurement can be performed in several facilities such as RIBLL[4] in HIRFL[5].

1. Chiara, C. J. *et al.* Isomer depletion as experimental evidence of nuclear excitation by electron capture. *Nature* **554**, 216 (2018).

2. Wu, Yuanbin, Keitel, Christoph H. & Pálffy, Adriana. $^{93m}$Mo isomer depletion via beam-based nuclear excitation by electron capture. *Phys. Rev. Lett.* **122**, 212501 (2019).

3. Fukuchi, T. *et al.* High-spin isomer in $^{93}$Mo. *Eur. Phys. J. A* **24**, 249-257 (2005).

4. Sun, Z. Y. et al. RIBLL, the radioactive ion beam line in Lanzhou. *Nucl. Instrum. Methods A* **503**, 496 (2003).

5. Xia, J. W. et al. The heavy ion cooler-storage-ring project (HIRFL-CSR) at Lanzhou. *Nucl. Instrum. Methods A* **488**, 11 (2003).

**Acknowledgements** This work was supported by the Strategic Priority Research Program of Chinese Academy of Sciences (Grant No. XDB34010000) and the National Natural Science Foundation of China, under contract Nos. U1832134 and U1932137.

**Author Contributions** S.G found the matter and wrote the first draft. Y.D.F, X.H.Z. and C.M.P edited and revised the manuscript.

**Supplementary Information**

We examined the chance coincidences between the 268-keV γ-ray and the transitions above the isomer, which were concluded to be negligible in Ref. 1. Based on the reported experimental setup, we plotted in Extended Data Fig. 1 the cross section of $^{93}$Mo as a function of bombarding energy in different layers. Note that this nucleus can be produced also from the reaction between the $^{90}$Zr beam and carbon layers. With a reported beam intensity of about $6 \times 10^8$ ions s$^{-1}$, the total production rate of $^{93}$Mo is estimated to be ~9.3 kHz. According to the calculation of PACE4[6] approximately 90% of the $^{93}$Mo residues were populated with spins larger than $17/2\hbar$. Approximately 20 - 30% of them will go through the 268-keV transition according to the relative intensities of 263-, 268-, 770-, 963-keV transitions estimated from Fig. 2b of Ref. 1, leading to a production rate ($R_{p268}$) on this transition of 1.7 – 2.5 kHz. Considering a coincidence time window ($T_w$) which should be between 100 ns and 1 μs according to Ref. 1, it can be recorded with any transition with no true relation with it with a probability of $P_{chance} = 2 \times R_{p268} \times T_w = 0.03 – 0.5\%$. Being neglected, it contributed to the deduced excitation probability (~1%).

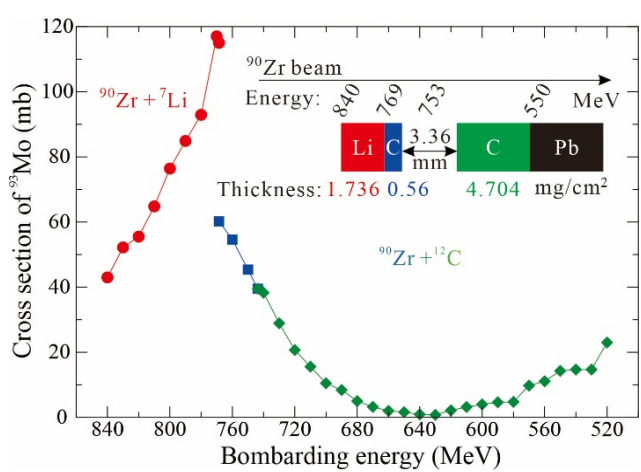

**Extended Data Figure 1 | Experiment setup and estimated cross-section of $^{93}$Mo as a function of bombarding energy.** In the upper right, we show the multi-layers on the path of the beam and the bombarding energies at the interfaces. The effective thicknesses of the layers are shown considering the 12% increase due to the angle with the beam path (see the methods part in Ref. 1). The cross-sections were calculated using the Monte Carlo fusion-evaporation code PACE4[6].

According to the Method section of Ref. 1, a ratio $k$, defined by $k = (g_1g_2-g_1b_2)/(b_1g_2-b_1b_2)$ is employed to subtract the background of doubly gated spectra, where a gate is regarded to be composed of the peak of interest (p) atop a smooth background (b), $g = p + b$. Here one transition is assumed to be in true coincidence with the transitions of interests, and marked by the subscript 2, while the other one is not in true coincidence with them without the isomer depletion mechanism, and marked by the subscript 1.

In the Extended Data Fig. 3 of Ref. 1, an example for deducing the $k$ ratios is provided. In the spectra of that figure, subscripts 1 and 2 correspond to the Doppler-shifted 2,475-keV and the unshifted 1,478-keV γ-rays, respectively. Four transitions at 123, 203, 770 and 963 keV were used to deduce the average ratio ($k_{ave}$). Since they together with the 268-keV transition, are not visible above the background in spectra $g_1b_2$ and $b_1b_2$ (see the Extended Data Figs. 3b and 3d of Ref. 1), the definition of $k$ can be simplified as $k = g_1g_2/b_1g_2$ for them, without a dramatic change on the central value.

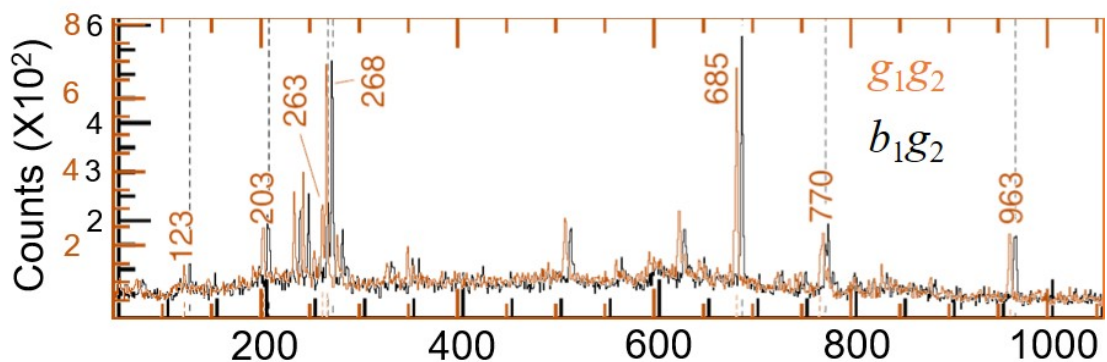

**Extended Data Figure 2 | Re-examination for the $k$ ratios.** This figure is obtained by overlapping the Extended Data Figs. 3a ($g_1g_2$ in orange) and 3c ($b_1g_2$ in black) of Ref. 1 using Microsoft PowerPoint. The y-axis of the spectrum in black color is extended so that the ordinates of the spectrum in orange color is 1.33 times that of the spectrum in black reaching thus the same heights. The spectrum in orange is slightly shifted to the left for easier comparison. The 770-keV peaks are located at the left shoulder of stronger 773-keV peaks.

However, according to a simple comparison plotted in Extended Data Fig. 2 of Ref. 1, the peak heights and background heights are similar for the five transitions in the $g_1g_2$ and $b_1g_2$, indicating that all $k$ ratios should be close to 1.33, or at least with a much large error. This estimated value is in agreement with the reported $k = 1.33(7)$ for the 268-keV transition, but seriously deviates from that of the 123-, 203-, 770- and 963-keV transitions which is $k_{ave} = 0.99(8)$. It is confusing that this value is not in coherent with the spectra to show its validity.

Furthermore, according to the definition of the $k$ ratios, it is not adequate to be used in obtaining the spectra shown in Fig. 3 of Ref. 1, since the transitions of interests are not in coincidence with both the two gating transitions without the isomer depletion mechanism. Therefore, the data analysis using $k$ ratios introduced in the Method section of Ref. 1, is adequate for the discussion of Fig. 2, but not for Fig. 3 in Ref. 1. The details to deduce the excitation probability and its error, was not described in Ref. 1.

6. Tarasov, O. B. & Bazin, D. Development of the LISE: application to fusion-evaporation. *Nucl. Instrum. Methods B* **204**, 174-178 (2003).